\documentstyle[twoside,fleqn,espcrc2]{article}

\def\sw2{\sin^2\theta_W}
\def\stw{\sin 2\theta_W}

\def\ct2w{\cot 2\theta_W}
\def\tb{\tan\beta}
\def\sq{\tilde q}

\def\slepton{\tilde\ell}

\def\ch#1{\omega_#1}

\def\xj{\chi_j}
\def\g{\tilde g}
\def\mgr{m_{3/2}}
\def\msq{\tilde {M_q}}
\def\mw{m_{\omega}}
\def\m#1{\tilde {m_#1}}
\def\mH{m_H}

\def\mw#1{m_{\omega #1}}

\def\mxi{m_{\chi i}}
\def\mxj{m_{\chi j}}

\def\mg{\tilde{m_3}}

\def\GL12{G_L^{12}}
\def\GR12{G_R^{12}}

\newcommand{\AmS}{{\protect\the\textfont2
  A\kern-.1667em\lower.5ex\hbox{M}\kern-.125emS}}

\title{
       \rightline{TKU-HEP 94/02}
       \rightline{IFM  2/94}
       $CP$ violation in minimal supersymmetric standard model
          \thanks{To appear in the proceedings of the Third KEK
                  Topical Conference on CP Violation, November 1993}
       }

\author{Y. Kizukuri\address{Department of Physics,
        Tokai University, \\
        1117 Kita-Kaname, Hiratsuka 259-12, Japan}%
        and
        N. Oshimo\address{Grupo Te\'orico de Altas Energias, CFMC, \\
        Av. Prof. Gama Pinto, 2, 1699 Lisboa Codex, Portugal}}

\begin{document}

\begin{abstract}
$CP$ violating phenomena predicted by the minimal supersymmetric
standard model are discussed in a case where the $CP$ violating
phases in SUSY sector are not suppressed.
The electric dipole moments of the neutron and the electron are large,
but can be smaller than their experimental upper bounds
if the scalar quarks and leptons are heavier than a few TeV.
$T$ violating asymmetries in the production processes of the different
neutralino pair and the different chargino pair emerge at the tree level.
They could be as large as of order $10^{-2}$
in unpolarized electron beam experiments
and $10^{-1}$ in polarized electron beam experiments.
In a pair production of the charginos of the same mass, the asymmetry emerges
through the electric and the weak "electric" dipole moments
of the charginos at the loop level, but its magnitude is at most
of order $10^{-4}$.

\end{abstract}

\maketitle

\section{INTRODUCTION}

Cabibbo-Kobayashi-Maskawa (CKM) matrix is a standard candidate
for the origins of $CP$ violation within a framework of the standard
electroweak theory.  In the minimal supersymmetric standard model (MSSM)
\cite{MSSM} there appear some complex parameters if the supersymmetry
(SUSY) is not an exact symmetry but is broken.
So in addition to the phase of CKM matrix we have other sources
of $CP$ violation in MSSM.  It is well-known that these new $CP$
violating phases cause large electric dipole moments (EDMs)
\cite{SusyEdm,KOPRD45,KOPRD46} which might give a denial
or a signal of the existence of SUSY.

Not only EDMs are the $CP$ violating phenomena expected in MSSM,
but also $T$ or $CP$--odd asymmetries in the production of SUSY particles,
such as neutralinos and charginos \cite{Kizukuri,Oshimo,KizuOshi}.
These asymmetries are caused by tree diagrams so that
they would not be too small to be tested by the future collider experiments
in which the SUSY particles might be eventually produced.
In this report we shall discuss first EDMs and show that
SUSY particles, especially squarks and sleptons, are as heavy as
a few TeV if the phases of the SUSY parameters are of order unity.
Secondly we shall discuss the $T$--odd asymmetries and show that
it is not impossible to observe the asymmetries as the manifestation of
$CP$ violation originating from SUSY.

Before closing this section, for the definiteness we show the MSSM Lagrangian,
\begin{eqnarray}
\lefteqn{    L = L_{kin} + L_{gauge} + L_F + L_S;} \nonumber \\
        L_F &=& [ (E^c Y_E L)H_1 + (D^c Y_D Q)H_1 \nonumber \\
             &&+ (U^c Y_UQ)H_2 + m_HH_1\times H_2 ]_F, \nonumber \\
      - L_S &=& (\tilde E^c\eta_E\tilde L)\tilde H_1
                + (\tilde D^c\eta_D\tilde Q)\tilde H_1 \nonumber \\
             && + (\tilde U^c\eta_U\tilde Q)\tilde H_2
                +  \tilde M^2_H\tilde H_1\times \tilde H_2  \nonumber \\
             && + {1\over 2}\sum_{i=1}^3{\tilde m}_i\lambda_i\lambda_i
                + \sum_{a,b}\tilde M_{ab}^2\phi_a^*\phi_b,
\label{eq:MSSM}
\end{eqnarray}
to fix our notation of the relevant parameters.
We assume the grand unification of the fundamental interactions,
and relate some of these parameters in (\ref{eq:MSSM}) each other as
\begin{eqnarray}
\tilde M_{ab}^2 &=& |\mgr|^2 \delta_{ab}, \nonumber  \\
\eta_f &=& A\mgr Y_f\ \ \ \ \ \ \  ( f = E, D, U ),
                                          \nonumber  \\
\tilde M_H^2 &=& B\mgr m_H,               \nonumber  \\
{\tilde m}_i &=& M_\lambda
               \ \ \ \ \ \ \ \ \ \ \ \ \  ( i = 1, 2, 3 ),
\label{eq:GutCondition}
\end{eqnarray}
to reduce the number of the parameters.
The parameters $A\mgr$, $B\mgr$, $m_H$,  and $M_\lambda$
in (\ref{eq:GutCondition}) are complex in general,
and can become sources of $CP$ violation.
However, all the phases are not physical, but two of them are physical.
So in this report we take $A\mgr$ and $m_H$ to be complex,
and $B\mgr$ and $M_\lambda$ to be real, for simplicity.


\section{ELECTRIC DIPOLE MOMENTS}

EDMs of the quarks ( leptons ) are given by the one-loop diagrams
by exchange of the squarks $\sq$ ( sleptons $\slepton$ ) and
the charginos $\omega_i$, the neutralinos $\xj$, or the gluinos $\g$ (not for
the lepton EDM ).  People think that the gluino loop gives the largest
contribution to EDMs of the quarks because the strong coupling constant
is much larger than the electroweak coupling constant.
Its typical estimate commonly made for the $u$-quark is like
\begin{eqnarray}
 d_u^G/e
        &\simeq & {2\alpha_s\over 9\pi \mg}{{\rm Im}(A\mgr)m_u
                    \over \mg^2}I'(\msq^2/\mg^2), \nonumber \\
        I(x) &=& {1\over (1-x)^2}(1 + x + {2x\over 1-x}{\rm ln}x).
\label{eq:edmg}
\end{eqnarray}
Assuming that the gluino mass ($\mg$) and the squark masses ($\msq$) are
100 GeV, we would get the $u$-quark EDM of $10^{-22} e\cdot$cm.  This is
much larger than the present experimental upper bound of the neutron EDM,
$10^{-25} e\cdot$cm \cite{edmnexp}, so we might have to assume that
the SUSY parameters are almost real.  However fortunately or unfortunately,
the SUSY particles such as squarks and gluinos have not yet been discovered,
and their masses are larger than 100 GeV.  Since, the heavier they are,
the smaller EDM becomes, even if the SUSY phases are of order unity,
sufficiently large SUSY particle masses can reduce EDM.  Thus
it is not a bad question to ask how heavy the SUSY particles should be
to reduce EDM to the value smaller than its experimental bound.

To answer this question we calculate the EDM of the neutron arising from
the gluino, chargino, and neutralino loops taking maximal $CP$ violating
phases.  The results are shown in Figs.~1--4.
\begin{figure}[htb]
\includegraphics{edmfig3.ps}
\vspace{6.0cm}
\caption{(i) The chargino, (ii) neutralino, and (iii) gluino contributions
to the neutron EDM for $\tb = 2$, $\m2 = |\mH| = 0.5 $TeV. }
\label{fig:edmcng}
\end{figure}

\begin{figure}[htb]
\includegraphics{edmfig2.ps}
\vspace{6.0cm}
\caption{The neutron EDM from the chargino contribution. The values
of the parameters are shown in Table~1, and $\tb = 2$
for (a) and $\tb = 10$ for (b).}
\label{fig:edmc}
\end{figure}

\begin{figure}[htb]
\includegraphics{edmfig6a.ps}
\vspace{6.0cm}
\caption{The neutron EDM as a function of $|\mH|$ for $\tb = 2$.
The gaugino mass $\m2$ is 1 TeV for (a) and 3 TeV for (b).
The squark mass $\msq$ is 0.2 TeV for (i) and 1 TeV for (ii).}
\label{fig:edmnmH}
\end{figure}

\begin{figure}[htb]
\includegraphics{edmfig6b.ps}
\vspace{6.0cm}
\caption{The neutron EDM as a function of $\m2$ for $\tb = 2$.
The higgsino mass $|\mH|$ is 1 TeV for (a) and 3 TeV for (b).
The squark mass $\msq$ is 0.2 TeV for (i) and 1 TeV for (ii).}
\label{fig:edmnm2}
\end{figure}

\begin{figure}[htb]
\includegraphics{edmfig7.ps}
\vspace{6.0cm}
\caption{The electron EDM from the chargino contribution. The parameters are
the same as in Fig.~\ref{fig:edmc}.}
\label{fig:edme}
\end{figure}

\begin{table}[hbt]
\caption{The values of $\m2$ and $|\mH|$ in Figs.~\ref{fig:edmc}
         and \ref{fig:edme}}
\label{tab:parameter}
\begin{tabular}{lcccc} 
\hline
              &  (i)  &  (ii) & (iii) & (iv)  \\
\hline
$\m2$   (TeV) & $0.2$ & $1.0$ & $0.2$ & $1.0$ \\
$|\mH|$ (TeV) & $0.2$ & $0.2$ & $1.0$ & $1.0$ \\
\hline
\end{tabular}
\end{table}

The Fig.~\ref{fig:edmcng} shows as a function of the squark mass
the absolute values of EDMs by the chargino, neutralino, and gluino loops
which contribute to the neutron EDM.  It should be noted
that EDM from the chargino is larger than EDM from the gluino in this case,
which would be contrary to the naive expectation.  The reason of this comes
mainly from the GUT mass relation $(\alpha_2/\alpha_3) \mg = \m2$ which we
assumed in the calculation. The large $SU(3)$ coupling $\alpha_3$,
which could make (\ref{eq:edmg}) large, leads to the large gluino mass,
which almost cancels the enhancement by the strong coupling.
In addtion to this, since the mixing between left--handed and right--handed
squarks, which is needed for the EDM operator by the gluino loop, is small,
the magnitude of the gluino contribution to the neutron EDM is suppressed
to be smaller than that of the chargino.

The Fig.~\ref{fig:edmc} shows the chargino contribution to the neutron
EDM for the different values of the parameters, $\tb$, $\m2$, and $|\mH|$,
as a function of the squark mass.  In this parameter region the chargino
contributes to the neutron EDM the most, and therefore its contribution
can be taken to be approximately equal to the whole neutron EDM.
 From the figure it can be seen that,
if the squark mass is larger than about 3 TeV for $\tb =2$ and
7 TeV for $\tb = 10$, the predicted value can be smaller than
the present experimental upper bound of $10^{-25}e\cdot$~cm,
even though the SUSY $CP$ phase is of order one.
This is the light gaugino case.

Can the squarks be lighter than 3 TeV?
Figs.~\ref{fig:edmnmH} and \ref{fig:edmnm2} try to answer the question, and
show the squarks can be really as light as 0.2 TeV if $\m2$ and $|\mH|$ are
heavier than about 2 TeV.  However, since EDM does not decrease
as rapidly as in Fig.~\ref{fig:edmc}
(even begins to increase when $|\mH|$ becomes large!),
the light squark possibility would be denied if the experimental bound
on the neutron EDM is reduced by one order of magnitude.

The electron EDM is shown in Fig.~\ref{fig:edme}.  Its present experimental
upper bound is $10^{-26} e\cdot$cm \cite{edmeexp}.  If the slepton is heavier
than about 1 TeV for $\tb = 2$ and 4 TeV for $\tb = 10$,
the predicted value can be smaller than the present experimental limit.

\section{T(CP)--ODD ASYMMETRY}
In the previous section it was realized that the present experimental
bounds on the neutron and electron EDMs do not exclude
the possibility that the SUSY $CP$ phases are of order unity.
What $CP$ violating phenomena will be expected to come out
from these new phases in high energy collider experiments?
One possibility is the $T$--odd asymmetry \cite{Kizukuri,Oshimo,KizuOshi}
in the production and decay processes of the SUSY particles.
In MSSM the asymmetry can appear at the tree level.  To be definite,
let us consider the production processes of two different charginos
mediated by $Z$--boson,
\begin{equation}
e^+e^- \rightarrow \ch2^+ \ch1^-,
\label{eq:epmw12}
\end{equation}
and two different neutralinos,
\begin{equation}
e^+e^- \rightarrow \chi_i  \xj.
\label{eq:epmxij}
\end{equation}
The charginos and the neutralinos are the mixed states of the
$SU(2)\times U(1)$ gauginos and the higgsinos.  To get them as
the mass eigenstates, we have to diagonalize their mass matrices
by unitary matrices which contain imaginary phases originating from
the complex SUSY parameters.  When we rewrite the gaugino and higgsino
couplings to $Z$ by the mass eigenstates through the unitary transformations,
there appear off-diagonal couplings of the charginos and neutralinos to $Z$.
Their coupling constants are made from the complex elements of
the unitary matrices.  Moreover their relative phases can not be absorbed,
and remain physical.  Thus these off-diagonal couplings to $Z$ cause
$T(CP)$ violation at the tree level, and the violation can be expected
to be large enough to be observed.

First we discuss the $T$--odd asymmetry in the chargino production process
(\ref{eq:epmw12}).  When the spin of $\ch2$, $s_2$, is summed,
but the spin of $\ch1$, $s_1$ (which will be chosen
such as it is perpendicular to the interaction plane), is not summed,
the cross section becomes in the CM system of $e^+e^-$ as
\begin{eqnarray}
\lefteqn{\sum_{s_2}d\sigma(e^+e^- \rightarrow {\ch2}^+ {\ch1}^-)} \nonumber \\
\lefteqn{= {G_F^2\over 2\pi}{p\over \sqrt S}
             {M_Z^4\over (S-M_Z^2)^2 + \Gamma_Z^2M_Z^2}\times}
                                                             \nonumber \\
\lefteqn{\ \times \bigl[(f_L^2 + f_R^2)(
              (|\GL12|^2 + |\GR12|^2)(E_1E_2 + p^2/3)}        \nonumber \\
\lefteqn{\    + 2\Re(\GL12{\GR12}^*) \mw1\mw2 )}              \nonumber \\
\lefteqn{\    + {\rm sign}(s_1)(f_L^2 - f_R^2)\Im(\GL12{\GR12}^*)
                {\pi\over 2}\mw2\ p \bigr],}
\label{eq:CharginoCrossSection}
\end{eqnarray}
where $p$ is the magnitude of the momentum of the chargino,
$E_{1, 2}$ are the energies of ${\ch1}_{ ,2}$,
$G_{L, R}^{12}$ are the coupling constants of the charginos to $Z$
defined as
\begin{eqnarray}
\lefteqn{J^Z_{\mu} = e\bar{\ch1}\gamma_{\mu}[G_L^{11}P_- + G_R^{11}P_+]\ch1}
                                                         \nonumber   \\
         && + e\bar{\ch2}\gamma_{\mu}[ G_L^{22}P_- + G_R^{22}P_+ ]\ch2
                                                         \nonumber   \\
         && + e\bar{\ch1}\gamma_{\mu}[ G_L^{12}P_- + G_R^{12}P_+ ]\ch2
                                                         \nonumber   \\
         && + e\bar{\ch2}\gamma_{\mu}[ G_L^{21}P_- + G_R^{21}P_+ ]\ch1,
\label{eq:CharginoNeutralCurrent}
\end{eqnarray}
$f_{L, R}$ are the coupling constants of the electron to $Z$,
and sign$(s_1)$ is defined by the three momentum of the incident electron,
${\bf p^-}$, and the three momentum of $\ch1$, ${\bf p}_1$, as
sign$(s_1)$ = sign(${\bf s}_1\cdot ({\bf p^-\times p}_1)$).

The term breaking $T$ invariance is the last term
in (\ref{eq:CharginoCrossSection}).  Because it depends on $s_1$,
we must observe at least a spin state of either chargino
to detect the $T$ violation effect. The positive sign of $s_1$
means a vector product ${\bf s}_1\cdot ({\bf p^-\times p}_1)$
is positive, and the  negative sign of $s_1$ means
${\bf s}_1\cdot ({\bf p^-\times p}_1) < 0$.
Since these two states are transformed to each other by time reversal,
the difference of the cross section between these two states
manifests $T$ violation at the first-order perturbation.
The asymmetry of these two states
\begin{eqnarray}
\lefteqn{  A_T = } \label{eq:CharginoAsymmetry}\\
\lefteqn{       {d\sigma({\bf s}_1\cdot ({\bf p^-\times p}_1)>0) -
         d\sigma({\bf s}_1\cdot ({\bf p^-\times p}_1)<0) \over
         d\sigma({\bf s}_1\cdot ({\bf p^-\times p}_1)>0) +
         d\sigma({\bf s}_1\cdot ({\bf p^-\times p}_1)<0)},} \nonumber
\end{eqnarray}
may be a good observable to quantify how large the $T$ violation is.
As a numerical example, we show the result of a parameter set of
$\tb = 2$, $\m2 = 200$ GeV, and $\mH = 200 {\rm e}^{i\pi/4}$ GeV,
which leads to the chargino masses 133, 275 GeV,
and the neutralino masses 83, 145, 203, 278 GeV.
The neutron and electron EDMs for this parameter set is shown
in Table~\ref{tab:edm} by taking the squark and slepton masses for 3 TeV.
Their values are smaller than the present experimental bounds
\cite{edmnexp,edmeexp}.
\begin{table}[hbt]
\caption{Electric dipole moments}
\label{tab:edm}
\begin{tabular}{lcc} 
\hline
            & Neutron & Electron \\
            & [$10^{-25} e\cdot$~cm] & [$10^{-27} e\cdot$~cm] \\
\hline
Gluino      & $ 0.2  $ & $ /    $     \\
Chargino    & $ -1.0 $ & $-4.2  $     \\
Neutralino  & $ 0.003$ & $ 0.04 $     \\
Exp.        & $ < 1.2$ & $-2.7\pm8.3$ \\
\hline
\end{tabular}
\end{table}

The calculated asymmetry is shown in Fig.~\ref{fig:CharginoAsymmetry}
as a function of the CM energy ($\sqrt S$).
\begin{figure}[htb]
\special{epsfile=tce.ps hscale=0.46 vscale=0.46 
          hoffset=-5 voffset=60}
\vspace{5.5cm}
\caption{The $T$--odd asymmetry in the chargino production.
         The solid line corresponds to unpolarized electron beam,
         and the dashed line corresponds to left-handedly polarized
         electron beam.}
\label{fig:CharginoAsymmetry}
\end{figure}
When the incident electron beam is unpolarized, the magnitude
of the asymmetry is smaller than $10^{-2}$ despite the tree level effect.
This is due to the almost pure-axial coupling of the electron to $Z$.
When the incident electron beam is polarized, the asymmetry is enhanced
by about 6 times as large as the unpolarized case.

Secondly we discuss the $T$--odd asymmetry in the neutralino production
process (\ref{eq:epmxij}).  As in the chargino production,
by summing the spin of $\xj$, $s_j$, but the spin of $\chi_i$, $s_i$, is
not summed,  we get the cross section in the CM system of $e^+e^-$ as
\begin{eqnarray}
\lefteqn{\sum_{s_j} d\sigma(e^+e^- \rightarrow \chi_i \xj)}  \nonumber \\
&=& {G_F^2 \over 4\pi}{p\over \sqrt S}
    {M_Z^4\over (S-M_Z^2)^2 + \Gamma_Z^2M_Z^2}|O_Z|^2\times  \nonumber \\
&\times& \bigl[(f_L^2 + f_R^2)(E_iE_j + p^2/3 - \cos 2\delta_Z \mxi \mxj)
                                                              \nonumber \\
&+& {\rm sign}(s_i)(f_L^2 - f_R^2){\pi\over 4}\sin 2\delta_Z \mxj
              \ p\bigr],
\label{eq:NeutralinoCrossSection}
\end{eqnarray}
where $p$ is the magnitude of the momentum of the neutralino,
$E_{i, j}$ are the energies of $\chi_{i, j}$,
and sign$(s_i)$ = sign(${\bf s}_i\cdot ({\bf p^-\times p}_i))$.
$O_Z$ is the coupling constant of the neutralinos to $Z$ defined as
\begin{eqnarray}
\lefteqn{{\cal L}_{\rm int}^Z = {\strut e\over \displaystyle \stw}\bar\chi_i
                    \gamma_{\mu}( O_Z P_- - O_Z^*P_+ )\xj Z^{\mu}; }
                                                           \nonumber \\
                    &&P_{\pm} = (1 \pm \gamma_5)/2,
\end{eqnarray}
and $\delta_Z = \arg (O_Z)$.  Again we have the $T$ violating term
in (\ref{eq:NeutralinoCrossSection}), and we can define an asymmetry
\begin{eqnarray}
\lefteqn{  A_T = } \label{eq:NeutralinoAsymmetry}\\
\lefteqn{    {d\sigma({\bf s}_i\cdot ({\bf p^-\times p}_i)>0) -
         d\sigma({\bf s}_i\cdot ({\bf p^-\times p}_i)<0) \over
         d\sigma({\bf s}_i\cdot ({\bf p^-\times p}_i)>0) +
         d\sigma({\bf s}_i\cdot ({\bf p^-\times p}_i)<0)},} \nonumber
\end{eqnarray}
which quantifies how large the $T$ violation is.  It should be noted that,
differently from the chargino production, the final state of
the process (\ref{eq:epmxij}) is an eigenstate of $CP$ because of
the neutralinos being the Majorana particles, so that the asymmetry
(\ref{eq:NeutralinoAsymmetry}) is a $CP$--odd asymmetry.  Thus a non-vanishing
value of (\ref{eq:NeutralinoAsymmetry}) directly means the violation of $CP$.
The calculated asymmetry is shown in Fig.~\ref{fig:NeutralinoAsymmetry}
as a function of $\sqrt S$.
\begin{figure}[htb]
\special{epsfile=tne.ps hscale=0.46 vscale=0.46 
          hoffset=-5 voffset=60}
\vspace{5.5cm}
\caption{The $T$--odd asymmetry in the neutralino production.
         The solid line corresponds to unpolarized electron beam,
         and the dashed line corresponds to left-handedly polarized beam.}
\label{fig:NeutralinoAsymmetry}
\end{figure}
The magnitude of the asymmetry is approximately the same as the chargino one.
We also have calculated an asymmetry of the lepton pair produced
by the decay of $\xj$, $A'_T$, which is more realistic than
(\ref{eq:CharginoAsymmetry}) and (\ref{eq:NeutralinoAsymmetry})
because we only have to measure the momenta of the final state leptons,
${\bf p}_D^{\pm}$, intead of the spin of $\xj$
being experimentally difficult to be measured.
In Fig.~\ref{fig:NeutralinoAsymmetry} the result in the left-handedly
polarized electron beam case is shown by dots.
The magnitude of $A'_T$ is the same order as $A_T$ near the threshold,
but it rapidly decreases as $\sqrt S$ becomes large.
This is because, as $\sqrt S$ becomes large, $E_i$ also becomes large
so that ${\bf p}_D^+$ and ${\bf p}_D^-$ come to orient
to the same direction.  Since the $T$ violation term is proportional to
${\bf p^-\cdot (p}_D^+\times {\bf p}_D^-)$, $A'_T$ decreases as $\sqrt S$
gets large.  Therefore, if one wishes to get a large asymmetry,
the asymmetry should be measured near the threshold.

\section{DISCUSSION}
In this report we have viewed $CP(T)$ violation in MSSM.
The present experimental bounds on the neutron and electron EDMs do not
immediately rule out the possibility that the imaginary phases
of the SUSY parameters have their natural values of ${\cal O}(1)$.
If this is the case, the $CP$ violation originating from MSSM
could lead to $T(CP)$--odd phenomena in high energy experiments.
If the charginos and neutralinos are produced in $e^+e^-$ collisions,
$T$--odd asymmetry would be observed in the angular distribution
of the final decay products.

In the $T(CP)$-odd asymmetries discussed here two different mass eigenstates
of the charginos or neutralinos are involved.
In the pair production of the charginos or neutralinos
of the same mass eigenstate, does the asymmetry appear?
The answer is 'yes' for the charginos.
The electric and the weak "electric" dipole moments
of the charginos, $D_{\omega}$, are generated at one-loop level,
which break $T$ invariance. These dipole moment terms give rise to
$T$--odd asymmetry in the same chargino pair-production processes.
However, from the dimensional grounds, $A_T$ in this case can be
roughly estimated as $A_T \sim {\sqrt S}D_{\omega}/e$.
The main contributions to $D_{\omega}$ are given by the loop diagrams
involving the top quark, the top squarks, the $W$ bosons, and/or
the Higgs bosons.  The top or stop contribution to $D_{\omega}$ is
proportional to the large top-quark mass.  However, within MSSM
the magnitude of $D_{\omega}$ is strictly restricted by the upper bounds of
the neutron and electron EDMs, and it would be at most $10^{-20} e\cdot$cm,
if $m_t \sim 150$ GeV and unless one of the scalar quarks is accidentally
much lighter than the other scalar quarks and leptons.
The contributions from the $W$ bosons, and the Higgs bosons are roughly
estimated to be also $10^{-20} e\cdot$cm.  Thus we could expect
in $e^+e^- \rightarrow \omega_i \bar{\omega_i}$
$A_T \sim 10^{-4}$ which would be too small for detection
in the future $e^+e^-$ experiments.

\end{document}